\documentstyle[12pt,psfig,epsf]{article}
\newcommand{\be}{\begin{equation}}
\newcommand{\ee}{\end{equation}}

\newcommand{\beqa}{\begin{eqnarray}}
\newcommand{\eeqa}{\end{eqnarray}}

\newcommand{\eqref}[1]{(\ref{#1})}

\def\boxit#1{\vbox{\hrule\hbox{\vrule\kern8pt
\vbox{\hbox{\kern8pt}\hbox{\vbox{#1}}\hbox{\kern8pt}}
\kern8pt\vrule}\hrule}}
\def\mathboxit#1{\vbox{\hrule\hbox{\vrule\kern8pt\vbox{\kern8pt
\hbox{$\displaystyle #1$}\kern8pt}\kern8pt\vrule}\hrule}}

\def\IB{\relax\hbox{$\inbar\kern-.3em{\rm B}$}}
\def\IC{\relax\hbox{$\inbar\kern-.3em{\rm C}$}}
\def\ID{\relax\hbox{$\inbar\kern-.3em{\rm D}$}}
\def\IE{\relax\hbox{$\inbar\kern-.3em{\rm E}$}}
\def\IF{\relax\hbox{$\inbar\kern-.3em{\rm F}$}}
\def\IG{\relax\hbox{$\inbar\kern-.3em{\rm G}$}}
\def\IGa{\relax\hbox{${\rm I}\kern-.18em\Gamma$}}
\def\IH{\relax{\rm I\kern-.18em H}}
\def\IK{\relax{\rm I\kern-.18em K}}
\def\IL{\relax{\rm I\kern-.18em L}}
\def\IP{\relax{\rm I\kern-.18em P}}
\def\IR{\relax{\rm I\kern-.18em R}}
\def\IZ{\relax\ifmmode\mathchoice
{\hbox{\cmss Z\kern-.4em Z}}{\hbox{\cmss Z\kern-.4em Z}}
{\lower.9pt\hbox{\cmsss Z\kern-.4em Z}} {\lower1.2pt\hbox{\cmsss
Z\kern-.4em Z}}\else{\cmss Z\kern-.4em Z}\fi}

\def\II{\relax{\rm I\kern-.18em I}}

\begin{document}

\hfill  NRCPS-HE-02-06

\vspace{5cm}
\begin{center}
{\LARGE  Conformal Invariant Strings \\
with\\
Extrinsic Curvature Action\\
}

\vspace{2cm}

{\sl G.K.Savvidy\footnote{email:~savvidy@inp.demokritos.gr}\\
National Research Center Demokritos,\\
Ag. Paraskevi, GR-15310 Athens, Hellenic Republic\\

}
\end{center}
\vspace{60pt}

\centerline{{\bf Abstract}}

\vspace{12pt}

\noindent We study a string theory
which is exclusively based on extrinsic curvature action.
It is a {\it tensionless} string theory  because the action
reduces to  perimeter for the flat Wilson loop.
We are able to solve and quantize this high-derivative nonlinear
two-dimensional conformal field theory.
The absence of conformal anomaly in quantum theory requires
that the space-time should be 13-dimensional. We have found that
all particles, with arbitrary large spin, are massless.
This pure massless spectrum is consistent with the tensionless character
of the theory and we speculate that it may describe unbroken
phase of standard string theory when $ \alpha^{'} \rightarrow \infty$.


\newpage

\pagestyle{plain}

A string model which is exclusively based on the concept of extrinsic
curvature was suggested in  \cite{geo}.
It describes random surfaces embedded in D-dimensional
spacetime  with the following action
\be\label{funaction}
S =m \cdot L= {m\over \pi} \int d^{2}\zeta
\sqrt{h}\sqrt{K^{ia}_{a}K^{ib}_{b}},
\ee
where $m$ has dimension of mass, $h_{ab}$ is the induced metric and
$K^{i}_{ab}$ is the second fundamental form (extrinsic curvature).
In the above theory extrinsic curvature term $alone$ should be considered as
fundamental action of the theory.
The dependence on the extrinsic curvature in  (\ref{funaction})
guarantees proportionality of the action to the length $ L $.
The last property makes the theory very close to
the Feynman path integral for a point-like relativistic particle
because when the surface degenerates into a single world line the
action  (\ref{funaction}) becomes  proportional to the length of
the world line
\be\label{limit}
S ~~ \rightarrow ~~ m \int ds
\ee
and the functional integral over surfaces
naturally transforms into the Feynman path integral for a point-like
relativistic particle\footnote{This theory is essentially different
in its geometrical meaning with the actions considered in \cite{polykov}
and there is no $area$ term in the action
as it was in previous studies \cite{polykov}.}.

To illustrate these concepts let us consider a surface in
a form of cylinder of radius $R$ and of length $T$.
We can easily compute the action: the term $K^{ia}_{a}K^{ib}_{b}$
is equal to $1/R^2$ and the action is equal to $2T + 2\pi R$ (the
last term is a contribution coming from the boundary of the cylinder).
When the radius of the
cylinder $R$ tends to zero, the cylinder shrinks to a world line
of length $T$, and the action becomes proportional
to the length of the degenerated surface.

This string theory is {\it tensionless} because for the flat Wilson
loop the action is equal to its perimeter $S=m(R+T)$, and at the
{\it classical level string tension is equal to zero}.
In the recent articles \cite{geo,manvelyan} the authors studied
quantum fluctuations at one-loop level and demonstrated that
quantum fluctuations generate an area term, that is a string tension.
Our aim now is to further study the symmetry properties of
this model, specifically its conformal properties in order
to treat this nonlinear system exactly.

I shall consider two different theories:
the original model A when the metric tensor
on the world sheet is induced by embedding
into the target space-time and the second model B when
the world sheet metric is considered as an independent field.
We develop technical tools allowing to solve the model B and
to quantize this high-derivative nonlinear two-dimensional
conformal field theory. The contribution
to the central charge from bosonic coordinates is
twice bigger than in the standard bosonic
string theory. The contribution of Faddeev-Popov ghost and an anti-ghost
fields to the central charge remains the same,
therefore the absence of conformal anomaly requires
that  the space-time should be 13-dimensional. The mode
expansion of the fundamental fields allows to demonstrate that
{\it all particles, with arbitrary large spin, are massless}.
This pure massless spectrum is consistent with the tensionless character
of the model B and we speculate that it may describe unbroken
phase of standard string theory \cite{gross} when
$ \alpha^{'} \rightarrow \infty$ and
$M^{2}_{n} = {1\over \alpha^{'}}(n-1) \rightarrow 0$.
Large amount of {\it zero norm states} is an indication of enhanced symmetry
in our system.
David Gross defined this limit as an unbroken phase of string theory
where one should observe enhanced symmetry \cite{gross}.
We hope that this approach will also
provide better understanding of the original  theory A
which probably describe the broken string phase.

As in \cite{manvelyan} we shall represent the gonihedric action
(\ref{funaction}) in a form
\begin{equation}\label{conaction}
S= {m\over\pi}\int d^{2}\zeta \sqrt{h}\sqrt{ \left(\Delta(h)
X_{\mu}\right)^{2}},
\end{equation}
here ~$h_{ab}=\partial_{a}X_{\mu}\partial_{b}X_{\mu}$ ~is the induced
metric,~$\Delta(h)= 1/\sqrt{h}~\partial_{a}\sqrt{h}h^{ab}
\partial_{b} $ ~is  Laplace operator and
$K^{ia}_{a}K^{ib}_{b}=\left(\Delta(h) X_{\mu}\right)^{2}$. The
second fundamental form $K$ is defined through the relations:
$
K^{i}_{ab}n_{\mu}^{i}=\partial_{a}\partial_{b}X_{\mu}-
\Gamma^{c}_{ab}\partial_{c}X_{\mu}=\nabla_{a}\partial_{b}X_{\mu},
$
$ n_{\mu}^{i}n_{\mu}^{j}=\delta_{ij},\quad n_{\mu}^{i}
\partial_{a}X_{\mu}=0,$
where $n_{\mu}^{i}$ are $D-2$ normals and $a,b=1,2;~
\mu=0,1,2,...,D-1;\eta^{\mu\nu} =(-+..+);~  i,j=1,..,D-2.$

Below I shall consider the model B which has the same
action (\ref{conaction}) but now it should be interpreted as a functional of
two independent field variables $X^{\mu}$ and $h_{ab}$,
that is, we shall consider it as
two-dimensional quantum gravity interacting with scalar fields $X^{\mu}$.
At this stage the model B is not connected any more with any embedding
into space-time.
The aim is to compute the energy momentum tensor of the scalar field
$X^{\mu}$ and to demonstrate that it is traceless. Thus this theory
is conformally invariant at the classical level
\footnote{It is not obvious at all  that the last model is equivalent to the
original one even at the classical level, and as we shall
see it is  not, but this study will help to spot some
common properties and useful formulas.}.

To get classical equations and to construct world
sheet energy-momentum tensor one should
compute variation of the action with respect to the metric
$h^{ab}$ which is defined as
\be
\delta S = - {1\over 2\pi} \int \sqrt{h} T_{ab} \delta h^{ab} d^{2}\zeta.
\ee
The variation of the action is
\beqa
\delta S = {m\over\pi} \delta \int \sqrt{h}~ \sqrt{(\triangle (h)
X^{\mu})^{2}}~ d^{2}\zeta  ~~~~~~~~~~~~~~~~~~~~~~~~~~ \nonumber\\
={m\over\pi} \int(-{1\over 2}) \sqrt{h}~h_{ab}~ \delta h^{ab}
\sqrt{(\triangle (h) X^{\mu})^{2}}~ d^{2}\zeta +
{m\over\pi} \int \sqrt{h}~{\triangle (h) X^{\mu} \over
\sqrt{(\triangle (h) X^{\mu})^{2}}} ~\delta
\triangle (h) X^{\mu} ~ d^{2}\zeta \label{variation}
\eeqa
and we can proceed computing the variation of the Laplace operator
\be\label{varlap}
\delta ~\triangle (h) X^{\mu} = {1\over 2} {1\over \sqrt{h}}
h_{ab} \delta h^{ab} \partial_c \sqrt{h} h^{cd} \partial_d X^{\mu}
+ {1\over \sqrt{h}} \partial_a ( \sqrt{h} \delta h^{ab}
- {1\over 2} h^{ab}\sqrt{h} h_{cd}~ \delta h^{cd}  )
\partial_b X^{\mu}.
\ee
The first term on r.h.s. of (\ref{variation}) will be cancelled by the
first term coming from  (\ref{varlap}), therefore
\beqa\label{variation1}
\delta S =  {m\over\pi} \int { \triangle (h) X^{\mu}    \over
\sqrt{(\triangle (h) X^{\mu})^{2}} }~
\partial_a \{( \sqrt{h} \delta h^{ab}
- {1\over 2} h^{ab}\sqrt{h} h_{cd}~ \delta h^{cd}  )
~\partial_b X^{\mu} \}
~ d^{2}\zeta .
\eeqa
Integrating by part we can extract the $T_{ab}$ in covariant form
\be\label{cons1}
T_{ab} =  \nabla_{\{a } \left( m{ \triangle (h) X^{\mu} \over
\sqrt{(\triangle (h) X^{\mu})^{2}}} \right) ~\nabla_{b\}} X^{\mu}
-  h_{ab}~h^{cd} ~\nabla_{c}~ \left(m { \triangle (h) X^{\mu} \over
\sqrt{(\triangle (h) X^{\mu})^{2}}}
\right)~\nabla_{d} X^{\mu}
\ee
where $\{a ~~~b\} $ denotes a symmetric sum.  It is easy to see
that the energy-momentum tensor is traceless:
\be\label{conformal}
h^{ab} T_{ab} =0,
\ee
thus we have interaction with conformally
invariant matter field $X^{\mu}$. The
equation of motion, in a given case the constraint equation, is
\be\label{cons2}
T_{ab} = -~{2 \pi \over   \sqrt{h}~}{\delta S \over \delta h^{ab}}~ =~ 0.
\ee
We can get the equations which follow from the variation
of the action over coordinates $X^{\mu}$ as well:
\be\label{massshell}
{\pi \over   \sqrt{h}~}{\delta S \over \delta X^{\mu}} =
\triangle (h)~\left(m {\triangle (h) X^{\mu} \over
\sqrt{(\triangle (h) X^{\mu})^{2}}} \right) =0.
\ee
We can make all these formulas more transparent introducing
the operator $\Pi^{\mu}$
$$
\Pi^{\mu} =m {\triangle (h) X^{\mu} \over
\sqrt{(\triangle (h) X^{\mu})^{2}}},
$$
it has a property very similar with the constraint equation
for a point-like relativistic particle:
\be\label{secondcon}
(III)~~~~~~~~~~~~~~~~~~~\Theta ~\equiv ~\Pi^{\mu}\Pi^{\mu}~ -~ m^{2}~=0,~
\ee
and allows to rewrite the action (\ref{conaction}) in the form
\be\label{equivaction}
S= {1\over\pi}\int d^{2}\zeta ~\sqrt{h}~\Pi^{\mu}~ \Delta(h) X_{\mu}.
\ee
The equations now look like
\be\label{motioneq}
(I)~~~~~~~~~~~~~~~~~~~~~~~~~\triangle (h)~\Pi^{\mu} =0
\ee
and
\be\label{tensor}
(II)~~~~~~T_{ab} = \nabla_{\{a }\Pi^{\mu}  ~\nabla_{b\}} X^{\mu}
~-~~h_{ab}~h^{cd} ~ \nabla_{c}~ \Pi^{\mu}~\nabla_{d} X^{\mu}~=~0.
\ee
One can check the covariant conservation of the
energy momentum tensor
$
\nabla^{a}~T_{ab} =0.
$
Equations (\ref{motioneq}), (\ref{tensor})  and (\ref{secondcon})
completely define the system. We have equation of motion (\ref{motioneq})
together with the constraint equations (\ref{tensor})
and (\ref{secondcon}). We should stress that in
addition to the "standard" constraint
equation (\ref{tensor}) we have a new type of constraint (\ref{secondcon}).

We can fix the conformal gauge $h_{ab}=\rho\eta_{ab}$ using
reparametrization invariance of the action to derive it in the form
(see (\ref{conaction}),(\ref{equivaction}))
\be\label{gaga}
\acute{S} = {m\over\pi} \int d^{2}\zeta \sqrt{\left(\partial^{2}
X \right)^{2}} ~=~{1\over\pi}\int d^{2}\zeta ~\Pi^{\mu}~\partial^{2} X^{\mu}.
\ee
In this gauge the equations of
motion are more simple
\be\label{confequa}
(I)~~~~~~~~~~~~\partial^{2}\left(m \frac{\partial^{2}X_{\mu}}
{\sqrt{\left(\partial^{2}X \right)^2}}\right) =0.
\ee
and they should be accompanied by the constraint equations (\ref{secondcon})
and (\ref{tensor})
\beqa\label{cons3}
(II)~~~T_{ab} &=&  ~\partial_{\{a } \left( m{ \partial^{2} X^{\mu} \over
\sqrt{(\partial^{2} X )^{2}}} \right) ~\partial_{b\}} X^{\mu}
-~ ~\eta_{ab}~\partial_{c}~ \left( m{ \partial^{2} X^{\mu} \over
\sqrt{(\partial^{2} X )^{2}}}
\right)~\partial^{c} X^{\mu} =0,\nonumber\\
(III)~~~\Pi^{\mu}\Pi^{\mu}&-& m^{2}~=0,~~~~~~~~~~~~~~~\Pi^{\mu} = m \frac{\partial^{2}X^{\mu}}
{\sqrt{\left(\partial^{2}X \right)^2}}.
\eeqa
In the light cone coordinates
$\zeta^{\pm}=\zeta^{0}\pm\zeta^{1}$ the conformal gauge
looks like:
$
h_{++}=h_{--}=0, \qquad h_{+-}={1\over 2}\rho,
$
and the constrains (\ref{cons3}) take the form
\beqa\label{cons9}
T_{++}= {1\over 2} ~(T_{00} + T_{01})=
{1\over 2}(\partial_{0}\Pi^{\mu} +\partial_{1}\Pi^{\mu})
(\partial_{0}X^{\mu} +\partial_{1}X^{\mu})= 2~\partial_{+}\Pi^{\mu}
\partial_{+}X^{\mu},\nonumber\\
T_{--}= {1\over 2} ~(T_{00} - T_{01})= {1\over 2}
(\partial_{0}\Pi^{\mu} - \partial_{1}\Pi^{\mu})
(\partial_{0}X^{\mu} - \partial_{1}X^{\mu})= 2~\partial_{-}\Pi^{\mu}
\partial_{-}X^{\mu},
\eeqa
the trace is equal to the $T_{+-}$ component
$
2 T_{+-}= T_{00} - T_{11}=0.
$
The conservation of the energy momentum tensor takes the form
$
\partial_{-}T_{++} = \partial_{+}T_{--}=0
$
and  requires that its components are analytic $T_{++}=T_{++}(\zeta^{+})$
and anti-analytic $~T_{--}=T_{--}(\zeta^{-})$ functions. Thus our
system has infinite number of conserved charges.
This residual symmetry can be easily   seen in gauge
fixed action (\ref{gaga}) written in light cone coordinates
$$
\acute{S} = {4m\over\pi} \int \sqrt{(\partial_{+}\partial_{-}
X^{\mu})^{2}}~ d\zeta^{+}d\zeta^{-},
$$
it is invariant under the transformations
$
\zeta^{+} = f({\tilde{\zeta}^{+}}),~~~\zeta^{-} =
g({\tilde{\zeta}^{-}})
$
where $f$ and $g$ are arbitrary functions.

The classical equation  is $ \partial^{2}~\Pi^{\mu}=0 $,
therefore $\Pi^{\mu}$ is a function of the form
\be
\Pi^{\mu} =m~{\partial^{2} X^{\mu} \over
\sqrt{(\partial^{2} X)^{2}}} = {1\over 2}(\Pi^{\mu}_{L}(\zeta^{+})
+ \Pi^{\mu}_{R}(\zeta^{-})).\label{seconconsol}
\ee
Taking the ratio
$
{\Pi^{\mu} \over \Pi^{0} } = {\partial^{2} X^{\mu}
\over \partial^{2} X^{0}}
$
one can find the second derivative $\partial^{2} X^{\mu}$
written in the form
$$
\partial^{2} X^{\mu} = {\Pi^{\mu} \over \Pi^{0} } \partial^{2} X^{0}=
\Pi^{\mu} ~4\Omega(\zeta^+ ,\zeta^-) = {1\over 2}~
[\Pi^{\mu}_{L}(\zeta^{+}) + \Pi^{\mu}_{R}(\zeta^{-})]~4\Omega(\zeta^+ ,\zeta^-),
$$
where $\Omega(\zeta^+ ,\zeta^-)$ is another arbitrary function
of $\zeta^+$ and $\zeta^-$. Thus the equation reduces to the following
one:
$
\partial_{+}~\partial_{-} ~X^{\mu} = {1\over 2}[\Pi^{\mu}_{L}(\zeta^{+}) +
\Pi^{\mu}_{R}(\zeta^{-})]~\Omega(\zeta^+ ,\zeta^-),
$
and $X^{\mu}$ is a sum of inhomogeneous and homogeneous solutions
\be\label{I}
X^{\mu}= \Psi^{\mu}(\zeta^+ ,\zeta^-) +
{1\over 2}[X^{\mu}_{L}(\zeta^+) + X^{\mu}_{R}(\zeta^-)],
\ee
where inhomogeneous solution is
\be\label{II}
\Psi^{\mu}(\zeta^+ ,\zeta^-)= {1\over 2}\int^{\zeta^+}_{0}
\int^{\zeta^-}_{0}[\Pi^{\mu}_{L}(\tilde{\zeta}^{+}) +
\Pi^{\mu}_{R}(\tilde{\zeta}^{-})]~\Omega(\tilde{\zeta}^+ ,\tilde{\zeta}^-)
d\tilde{\zeta}^+d\tilde{\zeta}^-
\ee
and $X^{\mu}_{L}(\zeta^+)$, $X^{\mu}_{R}(\zeta^-)$
are arbitrary functions of $\zeta^+$ and $\zeta^-$.
The constraints (\ref{cons9}) take the form
\beqa
T_{++}=2~\partial_{+}\Pi^{\mu} \partial_{+}X^{\mu} = {1\over 2}~
\dot{\Pi}^{\mu}_{L}(\zeta^{+})
~\{\int^{\zeta^-}_{0} [\Pi^{\mu}_{L}(\zeta^{+}) +
\Pi^{\mu}_{R}(\tilde{\zeta}^{-})]~\Omega(\zeta^+ ,\tilde{\zeta}^-)
d\tilde{\zeta}^- +\dot{X}^{\mu}_{L}(\zeta^+)\},\nonumber\\
T_{--}= 2~\partial_{-}\Pi^{\mu}\partial_{-}X^{\mu}= {1\over 2}~
\dot{\Pi}^{\mu}_{R}(\zeta^{-})
~\{\int^{\zeta^+}_{0} [\Pi^{\mu}_{L}(\tilde{\zeta}^{+}) +
\Pi^{\mu}_{R}(\zeta^{-})]~\Omega(\tilde{\zeta}^+ ,\zeta^-)
d\tilde{\zeta}^+ + \dot{X}^{\mu}_{R}(\zeta^-)\}. \nonumber
\eeqa
Taking derivatives of the constraint equation $\Theta =0 ~$(\ref{secondcon}),
and using our solution (\ref{seconconsol}) one can see that
\be\label{secformcons}
\Theta^{(1,0)} \equiv \dot{\Pi}^{\mu}_{L}(\zeta^{+})[\Pi^{\mu}_{L}(\zeta^{+}) +
\Pi^{\mu}_{R}(\zeta^{-})]=0,~~~~~~\Theta^{(0,1)} \equiv
\dot{\Pi}^{\mu}_{R}(\zeta^{-})
~[\Pi^{\mu}_{L}(\zeta^{+}) +
\Pi^{\mu}_{R}(\zeta^{-})] =0,
\ee
thus
\be
T_{++}={1 \over 2}~\dot{\Pi}^{\mu}_{L}(\zeta^{+})~\dot{X}^{\mu}_{L}(\zeta^+),~~~~~
T_{--}={1 \over 2}~\dot{\Pi}^{\mu}_{R}(\zeta^{-})~\dot{X}^{\mu}_{R}(\zeta^-)],
\ee
verifying the  fact they are indeed functions of only one light cone variable.
We shall derive new constraints differentiating (\ref{secformcons}),
in particular
\be\label{morediriv}
\Theta^{(1,1)} \equiv \dot{\Pi}^{\mu}_{L}(\zeta^{+})
\dot{\Pi}^{\mu}_{R}(\zeta^{-}) = 0.
\ee
This formula has clear interpretation: left and right
movers are not independent and should be normal to each other.

The action (\ref{gaga}) is invariant under the global symmetries
$\delta X^{\mu} =\Lambda^{\mu\nu}X_{\nu} + a^{\mu}$,
where $\Lambda^{\mu\nu}$ is a constant antisymmetric matrix, while
$a^{\mu}$ is a constant. The translation invariance of the action (\ref{gaga})
$\delta_{a} X^{\mu} = a^{\mu}$ results into the conserved momentum
current
\be
P^{\mu}_{a} = \partial_{a}\Pi^{\mu} =
\partial_{a}\left(m \frac{\partial^{2}X^{\mu}}
{\sqrt{\left(\partial^{2}X \right)^2}}\right),~~~~~~~\partial^{a}
P^{\mu}_{a} =0,~~~~P^{\mu} = \int P^{\mu}_0 d\zeta^1
\ee
and Lorentz transformation
$\delta_{\Lambda} X^{\mu} = \Lambda^{\mu\nu}X_{\nu}$ into angular momentum
current
\be
M^{\mu\nu}_{a} = X^{\mu}\partial_{a}\Pi^{\nu} - X^{\nu}\partial_{a}\Pi^{\mu}
+ \Pi^{\mu}\partial_{a}X^{\nu} - \Pi^{\nu}\partial_{a}X^{\mu},~~~~~~~\partial^{a}
M^{\mu\nu}_{a} =0,~~~~M^{\mu\nu} = \int M^{\mu\nu}_0 d\zeta^1 .
\ee
From its definition the momentum density,
$
P^{\mu}(\zeta^{0},\zeta^{1})
\equiv P^{\mu}_{0}(\zeta^{0},\zeta^{1}) =
\partial_{0}\Pi^{\mu}
$
is conjugate to $X^{\mu}(\zeta^{0},\zeta^{1})$ therefore
$
[X^{\mu}(\zeta^{0},\zeta^{1}),P^{\nu}(\zeta^{0},\zeta^{'1})]
=  i\eta^{\mu\nu}
\delta(\zeta^{1} - \zeta^{'1})
$
and one can deduce that the following commutation relations should hold:
\beqa\label{basiccommutaors}
[\partial_{+} X^{\mu}_{L}(\zeta),
\partial_{+}\Pi^{\nu}_{L}(\zeta^{'})]= 2\pi i~ \eta^{\mu\nu} \delta^{~'}
(\zeta  - \zeta^{'}),\nonumber\\
~[\partial_{-} X^{\mu}_{R}(\zeta),
\partial_{-}\Pi^{\nu}_{R}(\zeta^{'})]= 2\pi i~ \eta^{\mu\nu} \delta^{~'}
(\zeta  - \zeta^{'}),
\eeqa
with all others equal to zero: ~$[\partial_{\pm} X^{\mu}_{L,R}(\zeta),
\partial_{\pm}X^{\nu}_{L,R}(\zeta^{'})]=0$,~~$[\partial_{\pm}
\Pi^{\mu}_{L,R}(\zeta), \partial_{\pm}\Pi^{\nu}_{L,R}(\zeta^{'})]=0$
\footnote{Requiring that commutation
relations for the energy-momentum tensor $T_{++}$ and $T_{--}$ should
form the algebra of the two-dimensional conformal group one can
get the same form of basic commutators (\ref{basiccommutaors}).}.
To make these formulas more transparent to the reader let me use
the analogy with the ghosts $c^{\pm}$ and anti-ghost $b_{\pm\pm}$ fields
(or super-ghorts), indeed
the standard Faddeev-Popov action has the form $\int c~\partial_{\pm} b$ with
nonzero anti-commutator only between $c$ and $b$ fields.
Making use of these commutators one can get extended algebra of constraints
\beqa\label{standardalg}
[~~T_{++}(\zeta)~, ~~~T_{++}(\zeta^{'})~]&=&
i \pi(T_{++}(\zeta) + T_{++}(\zeta^{'}))~\delta^{~'}
(\zeta - \zeta^{'}),~~~~~~~ \nonumber\\
~[~~T_{++}(\zeta)~, ~\Theta(\zeta^{'},\zeta^{-})~]&=&
-i\pi ~\Theta^{(1,0)}(\zeta,\zeta^{-})~\delta (\zeta - \zeta^{'}),~~~~~~~ \nonumber\\
~[T_{++}(\zeta),\Theta^{(1,0)}(\zeta^{'},\zeta^{-})]&=&~ i\pi~
\Theta^{(1,0)}(\zeta,\zeta^{-})~
\delta^{~'} (\zeta - \zeta^{'}),\nonumber\\
~[T_{++}(\zeta),\Theta^{(0,1)}(\zeta^{'},\zeta^{-})]&=&~-{i\pi\over 2}
\Theta^{(1,1)}(\zeta,\zeta^{-})~
\delta(\zeta - \zeta^{'}),
\eeqa
and so on. Similar relations holds for $T_{--}$.
The high derivative operators have been defined as
(see (\ref{secondcon}),(\ref{secformcons}),(\ref{morediriv}))~
$\Theta = \Pi^2 -m^2,~\Theta^{(1,0)} =P^{\mu}_{L}
\Pi^{\mu},~\Theta^{(0,1)} = \Pi^{\mu} P^{\mu}_{R},~
~\Theta^{(1,1)} =P^{\mu}_{L}P^{\mu}_{R}$,
and so on ($2P^{\mu}_{0} = \partial_{+}\Pi^{\mu}_{L} +
\partial_{-}\Pi^{\mu}_{R}
= P^{\mu}_{L} + P^{\mu}_{R}$). They form an infinite abelian algebra
\be\label{nonstandard}
~[\Theta^{(n,m)}, \Theta^{(k,l)}]=0~~~~
\ee
and hint at the existence of symmetries (\ref{standardalg}),
(\ref{nonstandard}) higher than the
conformal algebra.

From (\ref{gaga}) we can deduce the propagator
$
<\Pi^{\mu}(\zeta)X^{\nu}(\tilde{\zeta})>$ $={\eta^{\mu\nu}\over 2}
ln (\vert \zeta - \tilde{\zeta} \vert \mu).
$
Using explicit form of the solution (\ref{I}) and (\ref{II})
we should have
\beqa
<[\Pi^{\mu}_{L}(\zeta^{+})+ \Pi^{\mu}_{R}(\zeta^{-})]
~[X^{\nu}_{L}(\tilde{\zeta}^{+}) +
X^{\nu}_{R}(\tilde{\zeta}^{-}) +~~~~~~~~~~~~~~~~~~~~~~~~~~~~~~~\nonumber\\
+\int^{\tilde{\zeta}^+}_{0}
\int^{\tilde{\zeta}^-}_{0}[\Pi^{\mu}_{L}(\tilde{\tilde{\zeta}}^{+}) +
\Pi^{\mu}_{R}(\tilde{\tilde{\zeta}}^{-})]~\Omega(\tilde{\tilde{\zeta}}^+ ,
\tilde{\tilde{\zeta}}^-)
d\tilde{\tilde{\zeta}}^+d\tilde{\tilde{\zeta}}^-]>\nonumber\\
=~\eta^{\mu\nu}~[ln [( \zeta^{-} - \tilde{\zeta}^{-})\mu)]
~ +~ln [( \zeta^{+} - \tilde{\zeta}^{+})\mu)]]. \nonumber
\eeqa
Because there are no correlations between
right - left moving modes of $\Pi$ field
and the right - left moving modes of $X$ field we shall get
$
<\Pi^{\mu}_{R}(\zeta^-)X^{\nu}_{R}(\tilde{\zeta^-})> =
~\eta^{\mu\nu}~ln [( \zeta^{-} - \tilde{\zeta}^{-})\mu)],\\
<\Pi^{\mu}_{L}(\zeta^+)X^{\nu}_{L}(\tilde{\zeta^+})> =
~ \eta^{\mu\nu}~ln [( \zeta^{+} - \tilde{\zeta}^{+})\mu)].~
$
Now we are in a position to compute the two point correlation function
of the energy momentum operator:
\beqa
<T~T_{++}(\zeta^{+}) ~T_{++}(\tilde{\zeta}^{+})  >~ =~
{1\over 4} <T:\dot{\Pi}^{\mu}_{L}(\zeta^{+}) \dot{X}^{\mu}_{L}(\zeta^{+}):
:\dot{\Pi}^{\nu}_{L}(\tilde{\zeta}^{+})
\dot{X}^{\nu}_{L}(\tilde{\zeta}^{+}):>~ \nonumber\\
=~{1\over 4}   <\dot{\Pi}^{\mu}_{L}(\zeta^{+})
\dot{X}^{\nu}_{L}(\tilde{\zeta}^{+}) >
<\dot{X}^{\mu}_{L}(\zeta^{+}) ~\dot{\Pi}^{\nu}_{L}
(\tilde{\zeta}^{+}) > ={1\over 4} ~{D \over (\zeta^{+} -
\tilde{\zeta}^{+})^4}.
\eeqa
The ghost contribution to the central charge remains the same as for the
standard bosonic string:
$
-{13\over 4} ~{1 \over (\zeta^{+} -
\tilde{\zeta}^{+})^4},
$
therefore the absence of conformal anomaly requires that the
space-time should be 13-dimensional
\be
D_c = 13.
\ee
This result can be qualitatively understood if one takes
into account the fact that the field equations here are of the forth order
and therefore we have two time more degrees of freedom than in the
standard bosonic string theory.

Let us now find mode expansion of different
operators in the general solution (\ref{I}),(\ref{II}).
The appropriate boundary condition for closed strings
is simply periodicity of the coordinates $X^{\mu}(\zeta^{0},\zeta^{1})=
X^{\mu}(\zeta^{0},\zeta^{1} + 2\pi)$. The arbitrary periodic
functions $X^{\mu}_{L}$ and $X^{\mu}_{R}$  can be written
as normal mode expansions
\beqa
X^{\mu}_{L} = x^{\mu} +
{1\over m}\pi^{\mu}\zeta^{+} +
\sum^{\infty}_{n=1}\sqrt{{2\over n m^2}}~ \{q^{\mu}_{1n}\sin(n\zeta^{+}) +
q^{\mu}_{2n}\cos(n\zeta^{+})\},\nonumber\\
X^{\mu}_{R} =   x^{\mu}
+  {1\over m}\pi^{\mu}\zeta^{-} +
\sum^{\infty}_{n=1}\sqrt{{2\over  n m^2}}~ \{\tilde{q}^{~\mu}_{1n}\sin(n\zeta^{-}) +
\tilde{q}^{~\mu}_{2n}\cos(n\zeta^{-})\}\nonumber
\eeqa
and in similar manner $\Pi^{\mu}$
\beqa
\Pi^{\mu}_{L}=   m e^{\mu} +
 k^{\mu}\zeta^{+} + \sum^{\infty}_{n=1}\sqrt{{2m^2\over n}}
\{ -p^{\mu}_{1n}\cos(n\zeta^{+}) + p^{\mu}_{2n}\sin(n\zeta^{+})\},\nonumber\\
\Pi^{\mu}_{R}=   m e^{\mu} +
k^{\mu}\zeta^{-} + \sum^{\infty}_{n=1}\sqrt{{2m^2\over n}}
\{-\tilde{p}^{\mu}_{1n}\cos(n\zeta^{-}) + \tilde{p}^{\mu}_{2n}\sin(n\zeta^{-})\},
\nonumber
\eeqa
therefore  for $\partial_{\pm}\Pi^{\mu}$ we shall get
\beqa
P^{\mu}_{L}=\partial_{+}\Pi^{\mu} =
k^{\mu} + \sum^{\infty}_{n=1}\sqrt{2nm^2}
\{p^{\mu}_{1n}\sin(n\zeta^{+}) + p^{\mu}_{2n}\cos(n\zeta^{+})\}\nonumber\\
P^{\mu}_{R}=\partial_{-}\Pi^{\mu} =
k^{\mu} + \sum^{\infty}_{n=1}\sqrt{2nm^2}
\{\tilde{p}^{\mu}_{1n}\sin(n\zeta^{-}) + \tilde{p}^{\mu}_{2n}\cos(n\zeta^{-})\}.
\eeqa
We shall consider the coordinates $q_{1n},q_{2m}$ as internal degrees of
freedom.
Substituting the mode expansion into the commutator
$
[X^{\mu}_{L,R}(\zeta^{\pm}),
P^{\nu}_{L,R}(\zeta^{'\pm})]= 2 \pi i\eta^{\mu\nu} \delta
(\zeta^{\pm} - \zeta^{'\pm})
$
gives
$$
[e^{\mu}, \pi^{\nu}]=[x^{\mu}, k^{\nu}] =  i\eta^{\mu\nu},~~~[q^{\mu}_{i n},
p^{\nu}_{j m}] =  i\eta^{\mu\nu}\delta_{ij}  \delta_{n m},
$$
where $i,j = 1,2$. We can now deduce a less trivial commutator
$
[\partial_{\pm}X^{\mu}_{R,L}(\zeta^{\pm}),
\Pi^{\nu}_{R,L}(\zeta^{'\pm})]=  -2 \pi i\eta^{\mu\nu} \delta
(\zeta^{\pm} - \zeta^{'\pm})
$
with all other commutators equal to zero.
If one introduces operators
\beqa
b^{\mu}_{n} ={q^{\mu}_{1n} - i q^{\mu}_{2n}\over \sqrt{2}},~~~
b^{+\mu}_{n} ={q^{\mu}_{1n} + i q^{\mu}_{2n}\over \sqrt{2}}~~~~
a^{\mu}_{n} ={p^{\mu}_{2n} + i p^{\mu}_{1n}\over \sqrt{2}},~~~
a^{+\mu}_{n} ={p^{\mu}_{2n} - i p^{\mu}_{1n}\over \sqrt{2}},\nonumber
\eeqa
then one can see that the only nonzero commutator is
\be
[a^{\mu}_{n},b^{+\mu}_{m}]= \eta^{\mu\nu} \delta_{nm}
\ee
and our basic fields will have expansion
\beqa
P^{\mu}_{L}=k^{\mu} + \sum^{\infty}_{n=1} \sqrt{nm^2}
(a^{\mu}_{n} e^{-i n\zeta^{+}} + a^{+\mu}_{n} e^{ i n\zeta^{+}}),  \nonumber\\
\partial_{+}X^{\mu}_{L}= {1\over m}\pi^{\mu}  +
\sum^{\infty}_{n=1} \sqrt{n\over  m^2}
(b^{\mu}_{n} e^{-i n\zeta^{+}} + b^{+\mu}_{n} e^{ i n\zeta^{+}}).
\eeqa
Let us now define the Fourier expansion of our constraint operators, they are
the standard Virasoro operator $L$ and our new operators $\Theta$
\be
L_{n}  = <e^{in\zeta^+} :P^{\mu}_{L}~\partial_{+}X^{\mu}_{L}: >,~~~~
\Theta_{n,l}  =  <e^{in\zeta^+ + il\zeta^-}<:  \Pi^{\mu}~\Pi^{\mu} -m^2  :>
\ee
together with the rest high derivative operators
\beqa
\Theta^{(1,0)}_{n,l} &=& <~e^{in\zeta^+ + il\zeta^-}
:P^{\mu}_{L}~\Pi^{\mu}:>,~~~~~~~~~
\Theta^{(0,1)}_{n,l} = <e^{in\zeta^+ +
il\zeta^-}:\Pi^{\mu}~P^{\mu}_{R}:>,\nonumber\\
\Theta^{(1,1)}_{n,l} &=& <e^{in\zeta^+ + il\zeta^-}
:P^{\mu}_{L}~ P^{\mu}_{R}:>~~~~....
\eeqa
and so on. In order to have explicit form of these operators it is useful
to introduce following oscillators
\beqa
\alpha^{\mu}_{n}&=& m \sqrt{n}~a^{\mu}_{n},
~~~ n > 0,~~~~~\alpha^{\mu}_{0}=k^{\mu};~~~
\beta^{\mu}_{n}= {1\over m} \sqrt{n}~b^{\mu}_{n},~~~
n > 0,~~~~~\beta^{\mu}_{0}=\pi^{\mu}/m  \nonumber\\
\alpha^{\mu}_{-n}&=& m \sqrt{n}~a^{+\mu}_{n},
~~~ n > 0,~~~~~~~~~~~~~~~~~~~
\beta^{\mu}_{-n}= {1\over m} \sqrt{n}~b^{+\mu}_{n},~~~
n > 0,
\eeqa
with nonzero commutator
\be
[\alpha^{\mu}_{n}, \beta^{\nu}_{k}]= n~\eta^{\mu\nu}\delta_{n+k,0}.
\ee
The fields are now represented in the form
$
P^{\mu}_{L}= \sum \alpha^{\mu}_{n} e^{-i n\zeta^{+}},~
\partial_{+}X^{\mu}_{L}=
\sum \beta^{\mu}_{n} e^{-i n\zeta^{+}}
$
and allow to represent the constraint operators in the form
\beqa
L_{n} &=&\sum_{l}:\alpha_{n-l} \cdot\beta_{l}:\nonumber\\
\Theta_{0,0} &=& m^2( e^{2} -1)
+ \sum_{n \neq 0}
{1\over 4 n^2}:(\alpha_{-n}~\alpha_{n} +
\tilde{\alpha}_{-n}~\tilde{\alpha}_{n}):\nonumber\\
\Theta_{n,0} &=& {im\over n}~e \cdot \alpha_{n}
-{1\over 4}\sum_{l \neq 0,n}
{1\over (n-l)l}:\alpha_{n-l}\cdot\alpha_{l}:\nonumber\\
\Theta_{0,n} &=&{im\over n}~e \cdot \tilde{\alpha}_{n}
-{1\over 4}\sum_{l \neq 0,n}
{1\over (n-l)l}:\tilde{\alpha}_{n-l}\cdot\tilde{\alpha}_{l}:\nonumber\\
\Theta_{n,l} &=& -{1\over 2 nl}:\alpha_{n}\cdot\tilde{\alpha}_{l}:
~~...
\eeqa
and so on.
Direct computation with the use of basic commutators allows to find
extended algebra of constrains (\ref{standardalg}) for Fourier components
\beqa
[L_n , L_m] &=& (l-n) L_{n+m} + {D\over 6}(n^3 -n)\delta_{n+m,0}\nonumber\\
~[L_n , \Theta_{m,k}] &=& (m+n) \Theta_{n+m,k}~~~~(J=0),...
\eeqa
and so on. Here J denotes the conformal spin of the corresponding operators.

The important fact which uniquely define the spectrum of this
string theory is the time dependence of the operator $\Pi^2$
$$
(\Pi^2 -m^2) = k^2 ~\zeta_{0}^2 + 2\{m e \cdot k ~
+ ~ k\cdot\Pi_{oscil} \} \zeta_{0} + \Pi^{2}_{oscil} +
2m e \cdot \Pi_{oscil}
+ m^2( e^2 -1).
$$
The first operator diverges quadratically with $\zeta_0$
and the second one linearly. Therefore in order to have
normalizable states in physical Hilbert space
one should impose corresponding constraints.
We are enforced to define the physical Hilbert space as
\be
k^2 ~ \Psi_{phys} =0,~~~e \cdot k ~\Psi_{phys} =0,
~~~k \cdot \alpha_{n}  ~\Psi_{phys} =0,~~~k \cdot
\tilde{\alpha}_{n} ~\Psi_{phys} =0,~~~~~n>0.
\ee
The first equation states that {\it all physical states with different spins
are massless}. This is consistent with tensionless character
of the theory. We should also  impose the rest of the constraints
\beqa\label{last}
L_0 \Psi_{phys} = \{k \cdot \pi  +m \sum^{\infty}_{n=1}
n(a^{+}_n b_n + b^{+}_n a_n)\}\Psi_{phys} &=&0\nonumber\\
L_n \Psi_{phys}&=&0,~~~~~n > 0\nonumber\\
\Theta_{0,0}\Psi_{phys} = \{( e^{2} -1) + \sum^{\infty}_{n=1}
{1\over 2n}(a^{+}_{n}~a_{n} +
\tilde{a}^{+}_{n}~\tilde{a}_{n}) \}\Psi_{phys} &=&0\nonumber\\
\Theta_{n,l}\Psi_{phys}&=&0,~~~~~n,l > 0~ ..........
\eeqa
and so on. For the ground state $a_n |k,e,0> = b_n|k,e,0>=0$ we
have $k^2 = 0,~ek =0,~e^2 = 1,$~ and~$ k \pi = k\partial_e |k,e,0> =0$
and therefore massless states of integer spin $j=0,1,2,..$
with momentum vector $k^{\mu}$ and polarization
tensor $e^{\{\mu_1} ....e^{\mu_j \} }$.
Equations (\ref{last}) should guarantee that there is no
ghost states in our physical subspace. One can see also
large amount of {\it zero norm states}, an indication of enhanced symmetry
in our system. One should study in great details this Hilbert space
in order to learn more about spin content of the theory
and to prove the absence of the negative norm states.
The details will be given elsewhere.

Despite the fact that this model is not equivalent to the
original one, it may represent an important limit of
standard bosonic string theory when $ \alpha^{'} \rightarrow \infty$.
Indeed in this limit all string states become massless
$M^{2}_{n} = {1\over \alpha^{'}}(n-1) \rightarrow 0$.
David Gross defined this limit as an unbroken phase of string theory
where one should observe enhanced symmetry \cite{gross}.

Below I shall return to
the original model A which contains many elements of the above
model. As we have seen we can not simply consider the action
(\ref{conaction}) as a functional of two independent
field variables $X$ and $h$ without losing connection with
the original system. In order to consider them as
independent variables and at the same time to have
equivalent theory at the classical level we have to introduce
additional  Lagrange multipliers $\lambda^{ab}$. The corresponding
action is
\be
S={m\over \pi} \int d^{2}\zeta \left\{\sqrt{h}~\sqrt{\left(\Delta(h)
X_{\mu}\right)^{2}}+\lambda^{ab}\left(\partial_{a}X^{\mu}\partial_{b}X^{\mu}-
h_{ab}\right)\right\}.
\ee
and the equations of motion are:
\beqa\label{acsel}
\triangle (h)~\left({\triangle (h) X^{\mu} \over
\sqrt{(\triangle (h) X^{\mu})^{2}}} \right)
-2~{1\over \sqrt{h}}\partial_{a}\left(
\lambda^{ab}\partial_{b}X^{\mu}\right)=0,~~~
\partial_{a}X^{\mu}\partial_{b}X^{\mu}-h_{ab}=0,\\
t_{ab} = ~T_{ab} + {m\over \sqrt{h}}~\lambda_{ab}=0,
\eeqa
where  we already  know $T_{ab}$ (\ref{cons1}),(\ref{tensor}).
For the trace of the energy momentum tensor we have:
\be
h^{ab} t_{ab} = {m\over \sqrt{h}}~h^{ab}\lambda_{ab},
\ee
where we have used the fact that $h^{ab} T_{ab} =0$.
If we substitute the expression for $\lambda$ from the last equation
into the first one we shall get unique equation for the
$X^{\mu}$:
\be
\triangle (h)~\left(m{\triangle (h) X^{\mu} \over
\sqrt{(\triangle (h) X^{\mu})^{2}}} \right)
+~{2\over \sqrt{h}}\partial_{a}\left(\sqrt{h}T^{ab}\partial_{b}X^{\mu}\right)=0,
\ee
where $h_{ab}=\partial_{a}X^{\mu}\partial_{b}X^{\mu}$. We can get the
same equation by direct variation of the original action (\ref{conaction})
$
\delta S  = \int \{ {\delta S \over \delta h_{ab}}
{\delta h_{ab} \over \delta X^{\mu}}  \delta X^{\mu} +
{\delta S \over \delta X^{\mu}} \delta X^{\mu}  \}d^{2}\zeta
$
and substituting the expressions for ${\delta S \over \delta h_{ab}}$
(\ref{cons2}) and ${\delta S \over \delta X^{\mu}}$ (\ref{massshell}).
Again we can fix the conformal gauge $h_{ab}=\rho\eta_{ab}$ using
reparametrization, in this gauge the equations of motion are :
\be\label{111}
\partial^{2}\frac{\partial^{2}X_{\mu}}
{\sqrt{\left(\partial^{2}X^{\nu}\right)^2}}
-2~\partial_{a}\left(\lambda^{ab}\partial_{b}X^{\mu}\right)=0,~~~
\partial_{a}X^{\mu}\partial_{b}X^{\mu}-\rho\eta_{ab}=0,~~~
\rho T_{ab} +m~\lambda_{ab} =0.
\ee
where $T_{ab}$ is given in (\ref{cons3}).
Comparing these equations with the ones in the previous
section (see equations (\ref{confequa}) and (\ref{cons3})) one can
be convinced once again  that these two systems are not equivalent
even at the classical level, but are  close enough.

The author wish to thank A.Nicolaidis and I.Bachas for
the discussion of the tensionless limit of standard
string theory
and E.Floratos, R.Manvelyan and A.Nichols for useful remarks. This work
was supported in part by the EEC Grant no. HPRN-CT-1999-00161.

\vfill
\end{document}